\documentclass[11pt]{article}

\usepackage[margin=1in]{geometry}
\usepackage[T1]{fontenc}
\usepackage[utf8]{inputenc}
\usepackage{lmodern}
\usepackage{microtype}
\usepackage{amsmath,amssymb,amsthm,mathtools}
\usepackage{booktabs}
\usepackage{array}
\usepackage{enumitem}
\usepackage[numbers,sort&compress]{natbib}
\usepackage[colorlinks=true,allcolors=blue]{hyperref}
\usepackage{cleveref}
\usepackage{setspace}
\usepackage{authblk}
\usepackage{tabularx}
\allowdisplaybreaks

\newtheorem{proposition}{Proposition}[section]

\newcommand{\E}{\mathbb{E}}
\newcommand{\Var}{\operatorname{Var}}
\newcommand{\Cov}{\operatorname{Cov}}
\newcommand{\cor}{\operatorname{cor}}

\title{A Regression-Based Framework for the ACF, PACF, Durbin–Levinson Recursion, and One-Step-Ahead Prediction
}
\author[1]{Kellen Gong}
\author[2]{Fang Li}

\affil[1]{Yale University}

\affil[2]{Department of Mathematical Sciences,
Indiana University Indianapolis}

\date{}

\begin{document}
\maketitle

\begin{abstract}
The autocorrelation function (ACF) and partial autocorrelation function (PACF)
are foundational tools for identifying autoregressive moving-average (ARMA)
models, yet they are often introduced in ways that appear disconnected from
the regression concepts students already know. This paper develops a unified,
regression-based instructional framework for the ACF, PACF, Durbin--Levinson
recursion, and one-step-ahead prediction for weakly stationary time series.
We show that the ACF is the coefficient from a simple linear regression of a
mean-zero stationary process on one of its lagged values, while the PACF is
both the coefficient of the newest predictor in an expanding multiple
regression and the corresponding partial correlation. Using partial regression,
we derive the Durbin--Levinson updates for the newly added coefficient, the
existing regression coefficients, and the prediction-error variance from
familiar ordinary least-squares principles. Worked MA(1) and AR(1) examples
show how the characteristic cutoff and tailing-off patterns of the ACF and
PACF emerge naturally from this regression perspective. The same recursive
regression coefficients also determine the optimal linear one-step-ahead
predictor. The resulting framework provides a coherent instructional pathway
from regression to model identification, recursive estimation, and prediction
and suggests practical ways to connect introductory regression and time series
courses.

\end{abstract}


\section{Introduction}

One of the central challenges in teaching time series analysis is helping
students connect new concepts to statistical ideas they already understand.
Students entering a first course in time series analysis typically have
completed a course in simple and multiple linear regression. They know that a
regression coefficient measures the linear relationship between a response and
a predictor and that, in multiple regression, a coefficient represents the
contribution of one predictor after accounting for the remaining predictors.
These ideas become part of students' statistical intuition long before they
encounter autoregressive moving-average (ARMA) models.

Despite this preparation, many students experience the partial autocorrelation
function (PACF) as one of the most abstract concepts in an introductory time
series course. Unlike the autocorrelation function (ACF), whose interpretation
as the correlation between observations separated by a fixed lag is relatively
intuitive, the PACF is frequently introduced through the Durbin--Levinson
recursion, matrix formulas, or determinant identities. Although mathematically
elegant, these presentations often appear disconnected from the regression
framework that students have already mastered. As a consequence, students may
perceive the PACF as an entirely new statistical object requiring a specialized
recursive algorithm rather than as a familiar regression coefficient.

A simple classroom experience illustrates this disconnect. Suppose students
are asked the following question during the first week of a time series course.

\begin{quote}
``If today's observation $X_t$ is regressed on yesterday's observation
$X_{t-1}$, what does the regression coefficient represent?''
\end{quote}

Students who have completed an introductory regression course usually answer
without hesitation that the coefficient measures the linear relationship
between the two variables. Next ask a second question.

\begin{quote}
``Suppose $X_t$ is regressed on both $X_{t-1}$ and $X_{t-2}$. What does the
coefficient of $X_{t-2}$ represent?''
\end{quote}

Again, students typically answer correctly: it measures the contribution of
the second lag after accounting for the first. Many are surprised to learn
that these familiar regression coefficients are closely connected to two of
the most important quantities in time series analysis. The first regression
coefficient is the population ACF at lag one, while the coefficient of the
newest lag in the second regression is the population PACF at lag two. More
generally, the ACF at lag $k$ is the coefficient from a simple regression of
$X_t$ on $X_{t-k}$, while the PACF at lag $k$ is the coefficient associated
with the newest predictor in a multiple regression of $X_t$ on its previous
$k$ observations.

This classroom observation motivates the central idea of the present paper.
Rather than introducing the ACF and PACF as fundamentally new concepts, we
propose viewing them as direct applications of ordinary least-squares
regression. From this perspective, the ACF is the slope in a simple regression
of a stationary process on one of its own lagged values, whereas the PACF is
the coefficient of the newest lag after the intermediate lags have been
accounted for. The Durbin--Levinson recursion then emerges from a sequence of
partial-regression updates, and optimal linear one-step-ahead prediction
follows from the same regression framework.

The purpose of this paper is not to introduce new theory for ARMA models but
rather to present an alternative instructional framework for teaching several
classical ideas in time series analysis. We show that the ACF, PACF,
Durbin--Levinson recursion, and one-step-ahead prediction can all be derived
from and interpreted through ordinary least-squares regression. Worked
examples for the MA(1) and AR(1) models demonstrate how the familiar cutoff
and tailing-off patterns of the ACF and PACF emerge naturally from this
viewpoint, and we conclude with practical suggestions for incorporating the
approach into an undergraduate or beginning graduate time series course. The
primary contribution of this paper is pedagogical rather than theoretical.

\subsection*{Why a Regression-Based Perspective Helps Students}

The instructional framework developed in this paper is motivated by a simple
observation: students entering a first course in time series analysis already
possess substantial knowledge of simple regression, multiple regression, and
partial correlation. Rather than asking students to learn a collection of
apparently unrelated concepts and algorithms, the proposed framework builds
directly on this prior knowledge.

This perspective reduces the number of genuinely new concepts students must
master and allows greater emphasis on statistical reasoning and model
interpretation. Students can understand the PACF through the familiar ideas
of multiple regression and partial correlation, the Durbin--Levinson recursion
through successive regression updates, and one-step-ahead prediction through
the same expanding regression framework. In this way, model identification,
recursive estimation, and prediction become connected applications of familiar
regression principles rather than separate computational topics.

Throughout the paper, we consider a mean-zero, weakly stationary time series
$\{X_t\}$; the mean-zero assumption is made for notational convenience and
entails no loss of generality after centering.

To emphasize the continuity between regression analysis and time series
analysis, the paper is organized to parallel the sequence in which students
typically learn regression. We first reinterpret the ACF as the coefficient
from a simple regression, then show that the PACF is the coefficient of the
newest predictor in an expanding multiple regression model. Partial regression
then provides a natural derivation of the Durbin--Levinson recursion, and the
same recursive regression framework produces the coefficients required for
one-step-ahead prediction.

\section{Why Students Find the PACF Difficult}

One of the recurring instructional challenges in an introductory time series
course is explaining why the PACF behaves differently from the ordinary ACF.
Whereas the ACF is naturally interpreted as the association between
observations separated by a fixed time lag, the PACF is often introduced
through the Durbin--Levinson recursion or matrix formulas involving
autocovariances. Although mathematically elegant, these presentations can
obscure the statistical meaning of the PACF and make it appear disconnected
from concepts students already know.

The key conceptual difficulty is that the PACF measures the relationship
between $X_t$ and $X_{t-k}$ after the linear effects of the intermediate lags
have been removed. Students who have studied multiple regression already
understand the closely related idea that a regression coefficient measures the
contribution of one predictor after accounting for the remaining predictors.
Likewise, partial correlation describes the association between two variables
after removing the linear effects of other variables. These familiar concepts
provide the natural bridge to the PACF.

Without this connection, students may regard the PACF as an entirely new
statistical object and focus on memorizing recursive formulas rather than
understanding what those formulas represent. A regression-first presentation
instead begins with familiar ideas from multiple regression and partial
correlation and then applies them to lagged observations.

The next sections develop this connection formally. We first review the
population regression identity underlying the ACF and then use multiple and
partial regression to establish the corresponding interpretation of the PACF.

\section{A Regression View of the Autocorrelation Function}
The regression interpretation of the autocorrelation function follows almost immediately from ordinary least squares. Rather than beginning with the classical definition of the ACF, we begin with a regression model that is already familiar to students.

\subsection{Simple Regression Revisited}
Consider two mean-zero random variables $X$ and $Y$. Since both variables are centered, there is no need to include an intercept: in a linear regression with an intercept, the optimal intercept is $E(Y)-a E(X)=0$. Thus, among all linear predictors of the form
\[
\widehat{Y}=aX,
\]
ordinary least squares chooses the coefficient that minimizes
\[
\E\!\left[(Y-aX)^2\right].
\]
The minimizing coefficient is
\begin{equation}
 a
 =
 \frac{\Cov(X,Y)}
      {\Var(X)}
 =
 \rho_{XY}
 \frac{\sigma_Y}{\sigma_X},
\label{eq:populationOLS}
\end{equation}
where $\rho_{XY}$ denotes the population correlation coefficient.

When $X$ and $Y$ have the same variance, Equation~\eqref{eq:populationOLS} simplifies considerably:
\[
a=\rho_{XY}.
\]
Thus, whenever predictor and response share a common variance, the regression coefficient equals the correlation coefficient.

This seemingly elementary observation is the key to understanding the autocorrelation function.

\subsection{The ACF as a Regression Coefficient}
Let $\{X_t\}$ denote a stationary mean-zero time series with
\[
\gamma(k)=\Cov(X_t,X_{t-k}).
\]
Then its ACF at lag $k$ is 
\[
\rho(k)=\frac{\gamma(k)}{\gamma(0)}.
\]
Because stationarity implies
\[
\Var(X_t)=\Var(X_{t-k})=\gamma(0),
\]
the predictor and response have identical variances. Consequently, regressing the current observation on one of its previous values,
\[
X_t=aX_{t-k}+\varepsilon_t,
\]
gives
\[
a=\frac{\Cov(X_t,X_{t-k})}{\Var(X_{t-k})}=\rho_{X_t X_{t-k}}\frac{\sigma_{X_t}}{\sigma_{X_{t-k}}}=\rho(k).
\]
Therefore, the autocorrelation function is exactly the regression coefficient obtained by predicting the current observation from one previous observation.

This interpretation immediately provides an intuitive explanation of what the ACF measures. The ACF quantifies the expected change in the best linear predictor of $X_t$ associated with a one-unit change in the lagged observation $X_{t-k}$. No new statistical machinery is required beyond ordinary least squares.

\subsection{Prediction Error}

Students are also familiar with the coefficient of determination as a measure
of the proportion of variation in the response explained by a regression
model. At the population level, for the simple regression

\[
X_t=aX_{t-k}+\varepsilon_t,
\]
the squared correlation $\rho^2(k)$ is the proportion of the variance of $X_t$ explained by its linear projection
on $X_{t-k}$. Consequently, the proportion of variance left unexplained is $ 1-\rho^2(k)$. Since stationarity gives $ \Var(X_t)=\gamma(0)$, the residual variance is therefore

\begin{equation}
\Var(\varepsilon_t)
=
\gamma(0)\bigl(1-\rho^2(k)\bigr).
\label{eq:predictionerror}
\end{equation}

Equation~\eqref{eq:predictionerror} is more than a routine regression result.
Later, when an additional lag is introduced into an expanding regression
model, the squared partial autocorrelation plays the same role: it measures
the proportion of the remaining variance explained by the newly added
predictor. This observation leads directly to the prediction-error recursion
in the Durbin--Levinson algorithm.



\section{A Regression View of the Partial Autocorrelation Function}
The regression interpretation of the ACF immediately suggests a natural question. If the ACF is simply the coefficient obtained by regressing the current observation on a single lagged observation, what changes when additional lags are included in the regression model? The answer leads directly to the PACF.

\subsection{Multiple Regression and Partial Regression}

To parallel the mean-zero time-series setting used throughout the paper,
let $Y,X_1,\ldots,X_p$ be mean-zero random variables. Students who have
completed an introductory regression course are familiar with the
interpretation of coefficients in the population multiple regression
\begin{equation}
Y
=
\beta_1X_1
+
\beta_2X_2
+\cdots+
\beta_pX_p
+
\varepsilon.
\label{eq:multiple}
\end{equation}
Unlike the coefficient in a simple regression, the coefficient $\beta_j$
measures the contribution of the predictor $X_j$ after accounting for the
remaining predictors. Thus, each coefficient represents the additional
linear contribution of one predictor beyond that explained by the others.

An equivalent perspective comes from partial regression. To isolate the
contribution of $X_p$, we remove the linear effects of
$X_1,\ldots,X_{p-1}$ from both the response $Y$ and the predictor $X_p$.
The relationship between the resulting residuals is characterized by the
Frisch--Waugh--Lovell theorem.

\begin{proposition}[Frisch--Waugh--Lovell]
\label{prop:fwl}
In the population multiple regression
\[
Y
=
\beta_1X_1
+\cdots+
\beta_pX_p
+
\varepsilon,
\]
the coefficient on $X_j$ is the same as the slope obtained by regressing the
residualized response on the residualized predictor, after the linear effects
of the remaining predictors have been removed from both.
\end{proposition}

The Frisch--Waugh--Lovell theorem provides the key link between multiple
regression and the PACF. In an expanding regression on lagged observations,
it allows the coefficient of the newest lag to be obtained by regressing the
residualized response on the residualized new predictor. This observation
leads directly to the regression interpretation of the PACF.

\subsection{The PACF Through Partial Regression}

Suppose that the current observation is predicted from its previous $k$
observations,
\begin{equation}
X_t
=
\phi_{k,1}X_{t-1}
+\cdots+
\phi_{k,k}X_{t-k}
+
\varepsilon_t^{(k)}.
\label{eq:pacfregression}
\end{equation}
The coefficient $\phi_{k,k}$ measures the additional contribution of lag $k$
after the shorter lags have been included.

To apply the partial-regression interpretation from Section~4.1, let
$\varepsilon_t^{(k-1)}$ be the residual from regressing $X_t$ on
$X_{t-1},\ldots,X_{t-k+1}$, and let $\eta_{t-k}^{(k-1)}$ be the residual
from regressing $X_{t-k}$ on the same intermediate lags. By the
Frisch--Waugh--Lovell theorem, $\phi_{k,k}$ is the slope in the simple
regression
\begin{equation}
\varepsilon_t^{(k-1)}
=
\phi_{k,k}\eta_{t-k}^{(k-1)}
+
\varepsilon_t^{(k)}.
\label{eq:pacfpartialregression}
\end{equation}

Thus,
\begin{equation}
\phi_{k,k}
=
\frac{
\Cov\!\left(
\varepsilon_t^{(k-1)},
\eta_{t-k}^{(k-1)}
\right)
}{
\Var\!\left(
\eta_{t-k}^{(k-1)}
\right)
}=\rho_{\varepsilon_t^{(k-1)}
\eta_{t-k}^{(k-1)}}\frac{\sigma_{\varepsilon_t^{(k-1)}}}{\sigma_{\eta_{t-k}^{(k-1)}}}.
\label{eq:pacfslopepartialregression}
\end{equation}

\subsection{From Partial Regression to Partial Correlation}

The population partial autocorrelation at lag $k$, denoted by $\alpha(k)$,
is the partial correlation between $X_t$ and $X_{t-k}$ after the linear
effects of the intermediate lags
$X_{t-1},\ldots,X_{t-k+1}$ have been removed from both variables.
Equivalently, it is the correlation between the corresponding residuals,
so
\[
\alpha(k)
=
\cor\!\left(
\varepsilon_t^{(k-1)},
\eta_{t-k}^{(k-1)}
\right)=\rho_{\varepsilon_t^{(k-1)}
\eta_{t-k}^{(k-1)}}.
\]

To see why this partial correlation agrees with the regression coefficient
$\phi_{k,k}$ from Section~4.2, note that under stationarity, the two residual
regressions have the same covariance structure up to reversal of the lag order
and therefore have the same residual variance:
\[
\Var\!\left(\varepsilon_t^{(k-1)}\right)
=
\Var\!\left(\eta_{t-k}^{(k-1)}\right)
=
\sigma_{k-1}^2,
\]
where $\sigma_{k-1}^2$ denotes their common residual variance after accounting
for the previous $k-1$ lags.

Therefore, by Equation~\eqref{eq:pacfslopepartialregression} , the
slope in Equation~\eqref{eq:pacfpartialregression} equals the correlation
between the two residuals:
\[
\phi_{k,k}
=\rho_{\varepsilon_t^{(k-1)}
\eta_{t-k}^{(k-1)}}.
\]
Hence,
\begin{equation}
\boxed{
\alpha(k)=\phi_{k,k}.
}
\label{eq:pacf}
\end{equation}
Thus, the PACF has two equivalent interpretations: it is the partial
correlation between $X_t$ and $X_{t-k}$ after removing the effects of the
intermediate lags and the coefficient of the newest predictor in the
corresponding expanding multiple regression.


\section{Reinterpreting the Durbin--Levinson Algorithm Through Partial Regression}
The previous section established that the PACF is the coefficient of the newest predictor in an expanding regression model and, equivalently, the correlation between two residuals obtained by removing the effects of the intermediate lags. This characterization immediately suggests a natural question. Suppose we already know the regression based on the first $k-1$ lags. How can we efficiently update the regression after adding one more lag?

From the perspective of ordinary least squares, this is not a time series problem at all. It is the familiar regression problem of determining how the regression coefficients and the residual variance change when a new predictor is added to an existing model. The classical Durbin--Levinson algorithm provides exactly this sequence of updates. In this section we show that every step of the algorithm follows directly from the partial-regression framework developed in the previous section.

\subsection{Step 1: The Existing Regression}
Suppose that the regression based on the first $k-1$ lagged observations has already been fitted,
\begin{equation}
X_t
=
\sum_{i=1}^{k-1}
\phi_{k-1,i}X_{t-i}
+
\varepsilon_t^{(k-1)},
\label{eq:kminus1}
\end{equation}
where
\[
\varepsilon_t^{(k-1)}
=
X_t-
\sum_{i=1}^{k-1}\phi_{k-1,i}X_{t-i}
\]
is the corresponding regression residual.

The residual $\varepsilon_t^{(k-1)}$ represents the portion of the current observation that cannot be explained by the first $k-1$ lagged observations. In regression terminology, it is the unexplained variation remaining after fitting the $(k-1)$-predictor model.

\subsection{Step 2: Residualizing the New Predictor}
The next step is to determine whether lag $k$ contributes additional predictive information.

To accomplish this, we regress the new predictor $X_{t-k}$ on exactly the same set of shorter lags,
\begin{equation}
X_{t-k}
=
\phi_{k-1,k-1}X_{t-1}
+
\phi_{k-1,k-2}X_{t-2}
+\cdots+
\phi_{k-1,1}X_{t-k+1}
+
\eta_{t-k}^{(k-1)},
\label{eq:newpredictor}
\end{equation}
where
\[
\eta_{t-k}^{(k-1)}
=
X_{t-k}
-
\sum_{i=1}^{k-1}\phi_{k-1,k-i}X_{t-i}.
\]
The coefficients appear in reverse order because the covariance structure of a stationary time series is invariant under time shifts. Consequently, the regression of the oldest lag on the intermediate lags has exactly the same covariance structure as the regression of the current observation on the same intermediate lags.

The residual $\eta_{t-k}^{(k-1)}$ represents the component of lag $k$ that is not linearly explained by the shorter lags.

\subsection{Step 3: Computing the New PACF Coefficient}

By the partial-regression result established in Section~4, the coefficient of
the newly added predictor in the full regression is obtained by regressing the
residualized response $\varepsilon_t^{(k-1)}$ on the residualized predictor
$\eta_{t-k}^{(k-1)}$. Hence,
\begin{equation}
\phi_{k,k}
=
\frac{
\Cov\!\left(
\varepsilon_t^{(k-1)},
\eta_{t-k}^{(k-1)}
\right)
}{
\Var\!\left(
\eta_{t-k}^{(k-1)}
\right)
}
=
\alpha(k).
\label{eq:partialcoefficient}
\end{equation}

The remaining task is to express the covariance in
Equation~\eqref{eq:partialcoefficient} in terms of the autocovariance
function. Using the residual expressions from the preceding two steps,
the numerator is
\[
\begin{aligned}
\Cov\!\left(
\varepsilon_t^{(k-1)},
\eta_{t-k}^{(k-1)}
\right)
&=
\Cov\!\left(
X_t-\sum_{i=1}^{k-1}\phi_{k-1,i}X_{t-i},
\,
X_{t-k}-\sum_{i=1}^{k-1}\phi_{k-1,k-i}X_{t-i}
\right)\\
&=
\gamma(k)
-
\sum_{i=1}^{k-1}
\phi_{k-1,i}\gamma(k-i),
\end{aligned}
\]
where the last equality follows from stationarity and the orthogonality of
the regression residuals to the intermediate lagged predictors.

As established in Section~4, the residualized predictor has variance
\[
\Var\!\left(
\eta_{t-k}^{(k-1)}
\right)
=
\sigma_{k-1}^2.
\]

Using the same regression-orthogonality argument, the common residual variance can also be expressed in terms of the
autocovariance function as
\[
\sigma_{k-1}^2
=
\gamma(0)
-
\sum_{i=1}^{k-1}
\phi_{k-1,i}\gamma(i).
\]

Substituting these expressions into
Equation~\eqref{eq:partialcoefficient} gives
\begin{equation}
\boxed{
\phi_{k,k}
=
\frac{
\gamma(k)
-
\displaystyle\sum_{i=1}^{k-1}
\phi_{k-1,i}\gamma(k-i)
}{
\sigma_{k-1}^2
}=
\frac{
\gamma(k)
-
\displaystyle\sum_{i=1}^{k-1}
\phi_{k-1,i}\gamma(k-i)
}{
\gamma(0)
-
\displaystyle\sum_{i=1}^{k-1}
\phi_{k-1,i}\gamma(i)
}.
}
\label{eq:newcoefficient}
\end{equation}
Dividing the numerator and denominator by $\gamma(0)$ gives the equivalent
autocorrelation form
\begin{equation}
\boxed{
\phi_{k,k}
=
\frac{
\rho(k)
-
\displaystyle\sum_{i=1}^{k-1}\phi_{k-1,i}\rho(k-i)
}{
1
-
\displaystyle\sum_{i=1}^{k-1}\phi_{k-1,i}\rho(i)
}.
}
\label{eq:newcoefficientacf}
\end{equation}

Thus, the classical Durbin--Levinson update for the new PACF coefficient
follows directly from the partial-regression representation developed in
Section~4.

\subsection{Step 5: Updating the existing regression coefficients}
Once the coefficient of the newly added predictor has been determined, the remaining regression coefficients must be adjusted to account for the additional explanatory variable.

Consider the expanded regression model,
\begin{equation}
X_t
=
\sum_{i=1}^{k-1}
\phi_{k,i}X_{t-i}
+
\phi_{k,k}X_{t-k}
+
\varepsilon_t^{(k)}.
\label{eq:fullregression}
\end{equation}
The coefficients $\phi_{k,i}$, $i=1,\dots,k-1$, are generally different from those obtained in the previous regression because the addition of a new predictor changes the least-squares solution.

To derive the update formula, substitute the regression of the new predictor, Equation~\eqref{eq:newpredictor}, into Equation~\eqref{eq:fullregression}:
\begin{align}
X_t
&=
\sum_{i=1}^{k-1}\phi_{k,i}X_{t-i}
+
\phi_{k,k}\left(
\sum_{i=1}^{k-1}\phi_{k-1,k-i}X_{t-i}
+
\eta_{t-k}^{(k-1)}
\right)
+
\varepsilon_t^{(k)} \nonumber\\
&=
\sum_{i=1}^{k-1}
\left(
\phi_{k,i}+\phi_{k,k}\phi_{k-1,k-i}
\right)X_{t-i}
+
\phi_{k,k}\eta_{t-k}^{(k-1)}
+
\varepsilon_t^{(k)}.
\label{eq:expandedmodel}
\end{align}

Using the partial-regression relation in
Equation~\eqref{eq:pacfpartialregression}, and comparing the coefficients of
the lagged predictors in Equation~\eqref{eq:expandedmodel} with those in the
$(k-1)$-lag regression, Equation~\eqref{eq:kminus1}, gives

\begin{equation}
\boxed{
\phi_{k,i}
=
\phi_{k-1,i}-\phi_{k,k}\phi_{k-1,k-i},
\qquad
 i=1,\ldots,k-1.
}
\label{eq:update}
\end{equation}
Equation~\eqref{eq:update} is the familiar coefficient update in the Durbin--Levinson recursion. From the regression perspective, it has a simple interpretation: adding a new predictor causes every existing coefficient to be adjusted in order to maintain the least-squares fit. Thus, the recursion is nothing more than the ordinary behavior of multiple regression after introducing an additional explanatory variable.

\subsection{Step 6: Updating the prediction error}
The final component of the Durbin--Levinson recursion follows directly from one of the most familiar identities in ordinary least squares.

Recall that the newly added predictor is introduced through the partial regression
\begin{equation}
\varepsilon_t^{(k-1)}
=
\phi_{k,k}\eta_{t-k}^{(k-1)}
+
\varepsilon_t^{(k)},
\label{eq:partialregression}
\end{equation}
where $\varepsilon_t^{(k-1)}$ is the residual from the regression based on the first $k-1$ lags, $\eta_{t-k}^{(k-1)}$ is the residualized new predictor, and $\varepsilon_t^{(k)}$ is the residual after including lag $k$.

Equation~\eqref{eq:partialregression} is a simple linear regression. Consequently, the
squared correlation

\[
\cor^2\!\left(\varepsilon_t^{(k-1)},\eta_{t-k}^{(k-1)}\right)
\]
represents the proportion of the remaining variance explained by the newly
added predictor. From Section~4, this correlation is exactly the population
partial autocorrelation, so the explained proportion is
\[
\phi_{k,k}^2.
\]
Students encounter the corresponding variance identity in every introductory
regression course:
\[
\boxed{\text{Residual Variance}=\text{Total Variance}\times(1-\phi_{k,k}^2).}
\]
In the present setting, the ``total variance'' is precisely the prediction-error variance from the regression using the first $k-1$ lags,
\[
\Var\!\left(\varepsilon_t^{(k-1)}\right)=\sigma_{k-1}^2,
\]
while the residual variance after introducing lag $k$ is
\[
\Var\!\left(\varepsilon_t^{(k)}\right)=\sigma_k^2.
\]
The regression variance identity therefore immediately gives
\begin{equation}
\boxed{
\sigma_k^2=\sigma_{k-1}^2\left(1-\phi_{k,k}^2\right).
}
\label{eq:varianceupdate}
\end{equation}
Equation~(15) has a direct regression interpretation. Before lag $k$ is
added, the unexplained variance is $\sigma_{k-1}^2$, and the squared partial
autocorrelation $\phi_{k,k}^2$ represents the proportion of this variance
explained by the newly added lag. Thus,
$\sigma_{k-1}^2\phi_{k,k}^2$ is the reduction in prediction-error variance,
while the proportion remaining unexplained is $1-\phi_{k,k}^2$. The resulting
prediction-error variance is therefore
$\sigma_{k-1}^2(1-\phi_{k,k}^2)$, which is precisely the
Durbin--Levinson prediction-error variance recursion.

This completes the regression interpretation of the Durbin--Levinson
recursion. The new PACF coefficient, the updates of the existing regression
coefficients, and the reduction in prediction-error variance all follow from
ordinary least-squares regression. Thus, the recursion can be understood
entirely through familiar regression principles, requiring no additional
time-series machinery beyond the stationarity assumptions already imposed.



\subsection{The Durbin--Levinson algorithm revisited}
Collecting the preceding results yields the classical Durbin--Levinson algorithm, now interpreted entirely through regression.

\bigskip
\noindent\textbf{Initialization ($k=1$).}
Fit the simple regression
\[
X_t=\phi_{1,1}X_{t-1}+\varepsilon_t^{(1)}.
\]
Because predictor and response have the same variance,
\[
\phi_{1,1}=\rho(1),
\]
and
\[
\sigma_1^2=\gamma(0)\bigl(1-\rho(1)^2\bigr).
\]

\bigskip
\noindent\textbf{Iteration ($k\ge 2$).}
For each additional lag,
\begin{enumerate}
\item Residualize the response and the newly added predictor with respect to
the intermediate lagged predictors.
\item Compute the coefficient of the residualized predictor,
\[
\phi_{k,k}=\frac{\gamma(k)-\sum_{i=1}^{k-1}\phi_{k-1,i}\gamma(k-i)}{\sigma_{k-1}^2}.
\]
Equivalently, when working directly with the autocorrelation function,
Equation~\eqref{eq:newcoefficientacf} may be used.
\item Update the existing regression coefficients,
\[
\phi_{k,i}=\phi_{k-1,i}-\phi_{k,k}\phi_{k-1,k-i},\qquad i=1,\ldots,k-1.
\]
\item Update the prediction-error variance,
\[
\sigma_k^2=\sigma_{k-1}^2\bigl(1-\phi_{k,k}^2\bigr).
\]
\end{enumerate}

Each step corresponds to a familiar operation from ordinary least squares: adding one predictor, estimating its coefficient, adjusting the previous coefficients, and computing the new residual variance.

\subsection*{Teaching Point}
The Durbin--Levinson algorithm is traditionally introduced as a recursive
procedure involving autocovariances and prediction errors.
The regression perspective presented here reveals a much simpler underlying
structure.

Each iteration consists of four familiar regression operations:
fit the current regression model,
residualize the newly added predictor,
estimate the coefficient of the residualized predictor,
and update the residual variance using the familiar identity

\[
\text{Residual Variance}
=
\text{Total Variance}(1-\rho^2).
\]
where $\rho$ denotes the correlation between the residualized response and
the residualized predictor.

Consequently,
every component of the Durbin--Levinson recursion has a direct interpretation
within ordinary least squares.
The PACF is both the coefficient of the newest predictor and the corresponding
partial correlation, the coefficient updates reflect the addition of a new
explanatory variable, and the prediction-error variance update follows from
the regression relationship between residual variance and squared correlation.

Throughout this development, the key steps rely on familiar ideas from
ordinary least-squares regression. This interpretation unifies model
identification, recursive estimation, and prediction within a single
regression framework. Students therefore encounter the Durbin--Levinson
algorithm not as a separate time-series technique, but as the familiar process
of repeatedly extending a multiple regression model by one predictor at a time.

\section{Building the PACF One Regression at a Time}

The previous section showed that the Durbin--Levinson algorithm can be
understood as a sequence of expanding multiple regressions. In particular,
Equations~\eqref{eq:newcoefficientacf} and~\eqref{eq:update} provide recursive
updates for the newly added PACF coefficient and the existing regression
coefficients.

The purpose of this section is to illustrate how these coefficient updates
operate in practice. Rather than treating the MA(1) and AR(1) processes as
isolated examples, we apply the recursion step by step to show how the
characteristic behavior of the PACF emerges naturally from successive
regression models.

\subsection{A Worked Example: Building the PACF for an MA(1) Process}

Consider the stationary MA(1) process

\[
X_t
=
\varepsilon_t
+
0.6\,\varepsilon_{t-1}, 
\]
where $\{\varepsilon_t\}$ is a white-noise process with mean zero and variance
$\sigma^2$.

The autocorrelation at lag one is

\[
\rho(1)
=\frac{\gamma(1)}{\gamma(0)}=\frac{\Cov(\varepsilon_t
+
0.6\,\varepsilon_{t-1},\,\varepsilon_{t-1}
+
0.6\,\varepsilon_{t-2})}{\Var(\varepsilon_t
+
0.6\,\varepsilon_{t-1})}=
\frac{0.6}{1+0.6^2}
=
0.441.
\]
For $k\ge2$, $X_t$ and $X_{t-k}$ contain no common white-noise terms, so

\[
\rho(k)=\cor(X_t, X_{t-k})=0,
\qquad
k\ge2.
\]

We now construct the PACF recursively using the regression updates derived
in Section~5. Throughout this example, numerical values are rounded to three decimal places.

\subsubsection*{Step 1: Fit the First Regression}

The first regression contains only one predictor, \[ X_t = \phi_{1,1}X_{t-1} + \varepsilon_t^{(1)}. \] As shown in Section~3, the simple-regression coefficient equals the autocorrelation at lag one. Hence, \[ \phi_{1,1} = \rho(1) = 0.441. \] Because there are no intermediate lags to remove, the ACF and PACF coincide at lag one. This coefficient provides the initialization for the recursive
updates that follow.

\subsubsection*{Step 2: Add One More Predictor}

We now extend the regression by introducing the second lag,
\[
X_t
=
\phi_{2,1}X_{t-1}
+
\phi_{2,2}X_{t-2}
+
\varepsilon_t^{(2)}.
\]
Using $\rho(2)=0$, Equation~\eqref{eq:newcoefficientacf} gives
\[
\phi_{2,2}
=
\frac{
\rho(2)-\phi_{1,1}\rho(1)
}{
1-\phi_{1,1}\rho(1)
}
=
-\frac{0.441^2}{1-0.441^2}
=
-0.241.
\]
Equation~\eqref{eq:update} then updates the first regression coefficient,
\[
\phi_{2,1}
=
\phi_{1,1}
-
\phi_{2,2}\phi_{1,1}=0.441-(-0.241)(0.441) = 0.547.
\]

Notice that although the autocorrelation at lag two is zero, the newly added
regression coefficient is not. The recursive regression therefore yields a nonzero coefficient for the newly added lag, corresponding to the PACF at lag two.

\subsubsection*{Step 3: Continue the Recursion}

Introducing a third lag gives

\[
X_t
=
\phi_{3,1}X_{t-1}
+
\phi_{3,2}X_{t-2}
+
\phi_{3,3}X_{t-3}
+
\varepsilon_t^{(3)}.
\]

Applying Equation~\eqref{eq:newcoefficientacf} at lag three gives
\[
\phi_{3,3}
=
\frac{
\rho(3)-\phi_{2,1}\rho(2)-\phi_{2,2}\rho(1)
}{
1-\phi_{2,1}\rho(1)-\phi_{2,2}\rho(2)
}.
\]
Since $\rho(2)=\rho(3)=0$,
\[
\phi_{3,3}
=
\frac{-(-0.241)(0.441)}
{1-(0.547)(0.441)}
= 0.140.
\]

The existing coefficients $\phi_{3,1}$ and $\phi_{3,2}$ are updated using
Equation~\eqref{eq:update}. Repeating the same procedure at subsequent lags
produces the successive PACF coefficients
\[
\phi_{1,1}=0.441,\qquad
\phi_{2,2}=-0.241,\qquad
\phi_{3,3}=0.140,\qquad
\phi_{4,4}=-0.083,\qquad
\phi_{5,5}=0.050,\qquad \ldots
\]
Although the ACF is zero beyond lag one, the recursive regressions continue
to produce nonzero PACF coefficients whose magnitudes gradually decrease.
Thus, the familiar tailing-off behavior of the MA(1) PACF emerges naturally
from the successive addition of lagged predictors.



\subsection{A Contrasting Example: Why the Recursion Stops for an AR(1) Process}

The MA(1) example illustrates how successive regression updates generate a
PACF that decays gradually. An AR(1) process provides a useful contrast
because the recursion terminates immediately after the first step.

Consider the stationary AR(1) process

\[
X_t
=
\phi X_{t-1}
+
\varepsilon_t,
\qquad
|\phi|<1,
\]
where $\{\varepsilon_t\}$ is a white-noise process with mean zero and variance
$\sigma^2$.

For a stationary AR(1) process, the autocovariances satisfy
\[
\gamma(k)=\Cov(X_t, X_{t-k})=\Cov(\phi X_{t-1}+\varepsilon_t,  X_{t-k})=\phi\,\gamma(k-1),
\qquad k\ge1,
\]
since $\varepsilon_t$ is uncorrelated with past observations. Dividing by
$\gamma(0)$ gives
\[
\rho(k)
=
\phi\,\rho(k-1).
\]
Since $\rho(0)=1$, repeated application yields
\[
\rho(k)=\phi^k,
\qquad k=1,2,\ldots.
\]

Thus, unlike the MA(1) process, whose ACF cuts off after lag one, the AR(1)
process has an ACF that tails off geometrically.

\subsubsection*{Step 1: Fit the First Regression}

The first regression is
\[
X_t
=
\phi_{1,1}X_{t-1}
+
\varepsilon_t^{(1)}.
\]

As shown in Section~3, the simple-regression coefficient equals the
autocorrelation at lag one. Since $\rho(1)=\phi$ for an AR(1) process,
\[
\phi_{1,1}
=
\rho(1)
=
\phi.
\]

Because there are no intermediate lags to remove, the ACF and PACF coincide
at lag one, just as in the MA(1) example. This coefficient provides the
initialization for the recursive updates that follow.

\subsubsection*{Step 2: Why the PACF Is Zero Beyond Lag One}
After fitting the first regression,
\[
X_t=\phi X_{t-1}+\varepsilon_t^{(1)},
\]
the residualized response is simply the white-noise error,
\[
\varepsilon_t^{(1)}=\varepsilon_t.
\]

At the next iteration, the lag-two predictor is residualized with respect to
$X_{t-1}$. Because $\varepsilon_t$ is uncorrelated with all past observations,
it is also uncorrelated with this residualized predictor. Hence the
partial-regression coefficient is zero:
\[
\phi_{2,2}=0.
\]
Adding the second lag therefore does not change the fitted regression, and the
residualized response remains
\[
\varepsilon_t^{(2)}=\varepsilon_t.
\]

The same argument applies at each subsequent iteration. If the residualized
response remains $\varepsilon_t$, then it is uncorrelated with the newly
residualized lagged predictor
\[
\eta_{t-k}^{(k-1)}
=
X_{t-k}
-
\sum_{i=1}^{k-1}
\phi_{k-1,k-i}X_{t-i},
\]
because this predictor is a linear combination of past observations.
Therefore,
\[
\Cov\!\left(
\varepsilon_t,
\eta_{t-k}^{(k-1)}
\right)
=
0,
\]
and hence by Equation~\eqref{eq:partialcoefficient}
\[
\phi_{k,k}=0,
\qquad k\ge2.
\]

Thus, each newly added lag has a zero partial-regression coefficient, so the
residualized response remains $\varepsilon_t$ throughout the recursion.
Consequently, the PACF cuts off after lag one, even though the ACF continues
to decay geometrically.

This behavior contrasts directly with the MA(1) example. For an MA(1)
process, successive regressions continue to produce nonzero partial-regression
coefficients whose magnitudes gradually decrease, whereas for an AR(1)
process, the first lag captures all linear dependence relevant for predicting
$X_t$, leaving zero partial-regression coefficients at subsequent lags.
Thus, the familiar cutoff and tailing-off patterns of the PACF emerge
naturally from the regression framework rather than as rules to be memorized.


\subsection{What the Two Examples Teach}

The MA(1) and AR(1) examples illustrate that the familiar identification
rules for the ACF and PACF are not isolated properties of particular time
series models. Instead, they arise naturally from the behavior of ordinary
least squares as additional lagged predictors are introduced into an expanding
regression model.

For the MA(1) process, the autocorrelation vanishes after lag one, so the ACF
cuts off immediately. Nevertheless, successive regressions continue to
produce nonzero partial-regression coefficients whose magnitudes gradually
decrease. Consequently, the PACF tails off.

The AR(1) process illustrates the complementary situation. Once $X_{t-1}$ is
included in the regression, the residualized response is the white-noise error
$\varepsilon_t$, which is uncorrelated with the residualized predictors formed
from lags beyond one. Their partial-regression coefficients are therefore zero,
and the PACF cuts off after lag one even though the ACF tails off
geometrically.

Table~\ref{tab:comparison} summarizes these contrasting behaviors from the
regression perspective.

\begin{table}[ht]
\centering
\caption{Regression interpretation of the classical ACF/PACF identification
rules.}
\label{tab:comparison}

\begin{tabularx}{\textwidth}{>{\bfseries}p{3.5cm}XX}
\toprule
 & MA(1) & AR(1) \\
\midrule

ACF &
Cuts off after lag one &
Tails off geometrically \\

PACF &
Tails off &
Cuts off after lag one \\

Regression &
Each added lag continues to modify the regression fit &
The first lag captures all remaining linear dependence \\

Recursive PACF &
Continues updating &
Becomes zero after lag one \\

\bottomrule
\end{tabularx}

\end{table}

The regression interpretation therefore provides more than an alternative
derivation of the PACF: it explains why the classical identification rules
hold. Rather than memorizing which function cuts off for which model, students
can understand these patterns as consequences of how ordinary least squares
behaves when lagged predictors are added sequentially to a regression model.




\section{One-Step-Ahead Prediction as Regression}

The same regression framework extends naturally to one-step-ahead prediction.
For the mean-zero weakly stationary process considered throughout the paper,
suppose that $X_1,\ldots,X_n$ have been observed. The population linear
regression of $X_{n+1}$ on the observed history is
\begin{equation}
X_{n+1}
=
\phi_{n,1}X_n
+
\phi_{n,2}X_{n-1}
+\cdots+
\phi_{n,n}X_1
+
\varepsilon_{n+1}^{(n)}.
\label{eq:forecast_decomp}
\end{equation}
The corresponding optimal linear one-step-ahead predictor is
\begin{equation}
\widehat X_{n+1\mid n}
=
\phi_{n,1}X_n
+
\phi_{n,2}X_{n-1}
+\cdots+
\phi_{n,n}X_1,
\label{eq:forecast_hat}
\end{equation}
with prediction-error variance
\begin{equation}
\sigma_n^2
=
\Var\!\left(
X_{n+1}-\widehat X_{n+1\mid n}
\right)
=
\Var\!\left(
\varepsilon_{n+1}^{(n)}
\right).
\label{eq:forecast_variance}
\end{equation}

For $n=1$, the predictor is initialized by the simple-regression result from
Section~3:
\[
\phi_{1,1}=\rho(1),
\qquad
\widehat X_{2\mid1}=\rho(1)X_1,
\]
with prediction-error variance
\[
\sigma_1^2
=
\gamma(0)\left(1-\phi^2_{1,1}\right).
\]
For $n\ge2$, the coefficients and prediction-error variance are updated
recursively using the results of Section~5. In the present prediction setting,
the new coefficient is
\begin{equation}
\phi_{n,n}
=
\frac{
\gamma(n)
-
\displaystyle\sum_{i=1}^{n-1}
\phi_{n-1,i}\gamma(n-i)
}{
\sigma_{n-1}^2
},
\label{eq:forecast_newcoefficient}
\end{equation}
the existing coefficients are updated according to
\begin{equation}
\phi_{n,i}
=
\phi_{n-1,i}
-
\phi_{n,n}\phi_{n-1,n-i},
\qquad
i=1,\ldots,n-1,
\label{eq:forecast_lower_update}
\end{equation}
and the prediction-error variance becomes
\begin{equation}
\sigma_n^2
=
\sigma_{n-1}^2
\left(1-\phi_{n,n}^2\right).
\label{eq:forecast_variance_update}
\end{equation}

These three updates are precisely
Equations~\eqref{eq:newcoefficient}, \eqref{eq:update}, and
\eqref{eq:varianceupdate}, respectively, with $k=n$.

Thus, one-step-ahead prediction uses the same three recursive operations as
the PACF construction: compute the coefficient of the newly added lag, update
the existing coefficients, and update the prediction-error variance. The
diagonal coefficients $\phi_{k,k}$ form the PACF sequence used for model
identification, whereas the full set
$\phi_{n,1},\ldots,\phi_{n,n}$ determines the optimal linear predictor in
Equation~\eqref{eq:forecast_hat}. From the regression perspective, model
identification and prediction therefore arise from the same sequence of
partial regressions on an expanding set of lagged predictors.


\section{Instructional Implications}

The regression-based framework developed in this paper is intended to
complement rather than replace the traditional presentation of ARMA model
identification. Its principal goal is to help students recognize that many
concepts introduced in a first time series course are natural extensions of
ideas they already know from regression analysis.

Students may initially view the ACF, PACF, the Durbin--Levinson algorithm,
and forecasting as separate topics, each requiring its own definitions and
computational procedures. The regression perspective instead places these
ideas within a common framework: the ACF is the coefficient from a simple
regression, the PACF is the coefficient of the newest predictor in an
expanding multiple regression, the Durbin--Levinson algorithm consists of
successive partial-regression updates, and one-step-ahead prediction uses the
same recursively generated regression coefficients.

This perspective suggests several practical teaching strategies.

\subsection{Building on Prior Knowledge}

Because most students encounter regression before time series analysis,
instructors can introduce the ACF and PACF through regression models that
students already know how to interpret.

For example, students can first consider the simple regression
\[
X_t
=
\phi_{1,1}X_{t-1}
+
\varepsilon_t^{(1)},
\]
and then extend it to
\[
X_t
=
\phi_{2,1}X_{t-1}
+
\phi_{2,2}X_{t-2}
+
\varepsilon_t^{(2)}.
\]
The transition from simple to multiple regression naturally introduces the
distinction between autocorrelation and partial autocorrelation: the first
describes the linear association at a given lag, whereas the second measures
the additional linear contribution of a new lag after the intermediate lags
have been accounted for.

\subsection{Emphasizing Successive Regression Models}

Rather than presenting the Durbin--Levinson algorithm as a recursive
computation to memorize, instructors can emphasize that each iteration adds
one new lagged predictor to an existing regression model. Students already
know that adding a predictor may change the existing regression coefficients,
explain part of the remaining variation, and reduce the residual variance.
The Durbin--Levinson recursion applies these familiar regression ideas
sequentially to lagged observations.

In this way, the recursive formulas acquire direct statistical
interpretations. The newly added coefficient is the corresponding PACF
coefficient, the existing coefficients are adjusted to account for the new
predictor, and the prediction-error variance decreases according to the
proportion of remaining variation explained by that predictor.

\subsection{Using Worked Examples}

The MA(1) and AR(1) examples in Section~6 provide complementary illustrations
of this regression perspective. For the MA(1) process, successive regressions
continue to produce nonzero partial-regression coefficients whose magnitudes
gradually decrease, explaining why the PACF tails off.

For the AR(1) process, once the first lag has been included, the residualized
response is the white-noise error. Because this error is uncorrelated with
the residualized predictors formed from subsequent lags, their
partial-regression coefficients are zero. The PACF therefore cuts off after
lag one.

These examples encourage students to explain the characteristic ACF and PACF
patterns through familiar regression reasoning rather than treating the
cutoff and tailing-off properties as rules to be memorized.

\section{Discussion and Extensions}

The regression-based perspective developed in this paper offers several
advantages for introductory time series instruction.

First, it reduces the number of genuinely new concepts that students must
learn. Rather than viewing autocorrelation, partial autocorrelation, recursive
estimation, and forecasting as unrelated topics, students encounter them as
natural extensions of simple regression, multiple regression, partial
regression, and prediction.

Second, the regression interpretation provides a conceptual explanation for
the characteristic behaviors of the ACF and PACF. The familiar cutoff and
tailing-off patterns emerge from the successive addition of lagged predictors
to an expanding regression model rather than appearing as properties that
must simply be memorized.

Third, the framework naturally connects model identification and prediction.
The same regression recursion that generates the PACF also produces the
coefficients required for optimal linear one-step-ahead prediction. Students
can therefore view prediction as another application of the same regression
framework rather than as a separate computational topic.

The framework is intended primarily for courses in which students have
already completed an introductory regression course. In that setting,
instructors can explicitly connect previously learned concepts to new material
in time series analysis, emphasizing continuity across the statistics
curriculum and potentially reducing the number of unfamiliar ideas introduced
at once.

Although this paper focuses on stationary ARMA models, the same instructional
philosophy may extend to more advanced topics, including autoregressive
integrated moving-average (ARIMA) models, seasonal time series models, and
vector autoregressive (VAR) models. Developing regression-based presentations
of these topics provides a natural direction for future instructional work.

\section{Conclusion}

This paper has presented a regression-based framework for teaching four
foundational ideas in introductory time series analysis: the autocorrelation
function, the partial autocorrelation function, the Durbin--Levinson
algorithm, and one-step-ahead prediction.

The central observation is that these concepts can be understood through
familiar regression principles. The ACF is the coefficient from a simple
regression of a stationary process on one of its lagged values, while the
PACF is the coefficient of the newest predictor in an expanding multiple
regression model and, equivalently, the corresponding partial correlation.
The Durbin--Levinson algorithm follows from successive partial-regression
updates, with the prediction-error variance recursion arising from the
regression relationship between explained and residual variance. The same
recursively generated coefficients then determine the optimal linear
one-step-ahead predictor.

Viewed in this way, model identification, recursive estimation, and prediction
form a single conceptual progression rather than a collection of separate
procedures. By building directly on students' prior knowledge of regression,
the framework provides a coherent pathway into time series analysis and
strengthens the connection between regression analysis and ARMA modeling.



\end{document}